# Poverty mapping in Mongolia with AI-based Ger detection reveals urban slums persist after the COVID-19 pandemic

Jeasurk Yang[1 *], Sumin Lee[2], Sungwon Park[2], Minjun Lee[2], Meeyoung Cha[1 2 *]

[1]Max Planck Institute for Security and Privacy, Bochum, Germany
[2]School of Computing, Korea Advanced Institute of Science and Technology, Daejeon, Republic of Korea

**Correspondence:**
Jeasurk Yang (jeasurk.yang@mpi-sp.org)
Max Planck Institute for Security and Privacy
Universitätsstraße 140, 44799 Bochum, Nordrhein-Westfalen, Germany

Meeyoung Cha (mia.cha@mpi-sp.org)
Max Planck Institute for Security and Privacy
Universitätsstraße 140, 44799 Bochum, Nordrhein-Westfalen, Germany

**Abstract**
Mongolia is among the countries undergoing rapid urbanization, and its temporary nomadic dwellings—known as *Ger*—have expanded into urban areas. Ger settlements in cities are increasingly recognized as slums by their socio-economic deprivation. The distinctive circular, tent-like shape of gers enables their detection through very-high-resolution satellite imagery. We develop a computer vision algorithm to detect gers in Ulaanbaatar, the capital of Mongolia, utilizing satellite images collected from 2015 to 2023. Results reveal that ger settlements have been displaced towards the capital's peripheral areas. The predicted slum ratio based on our results exhibits a significant correlation (r = 0.84) with the World Bank's district-level poverty data. Our nationwide extrapolation suggests that slums may continue to take up one-fifth of the population after the COVID-19 pandemic, contrary to other official predictions that anticipated a decline. We discuss the potential of machine learning on satellite imagery in providing insights into urbanization patterns and monitoring the Sustainable Development Goals.

**Highlights**

- An AI model detects ger settlements, a specific form of slums in Mongolia, from very-high-resolution satellite images.
- Ger settlements have expanded to the periphery of the capital since 2015, similar to the urban sprawl phenomenon.
- Prediction shows a significant correlation (r = 0.84) with poverty statistics, validating an AI-based method.
- Nationwide extrapolation suggests that slum ratio persist at 21% in Mongolia from 2020, supplementing official reports.

## 1. Introduction

After the collapse of the Soviet Union and the socialist system in the 1990s, Mongolia underwent significant economic transformations (Park et al., 2017). These changes triggered a substantial influx of rural residents to Ulaanbaatar, the capital city (Xu et al., 2021). This population surge, coupled with a housing shortage, led to the emergence of extensive informal settlements on the city's outskirts (Park et al., 2019). These settlements expanded with *gers*—traditional white, circular tent-like dwellings used by nomadic herders on the Mongolian steppes—reflecting the cultural roots of rural migrants (Byambadorj et al., 2011). Ger settlements in urban areas have become more permanent, evolving into what are known as 'ger districts' (Hamiduddin and Plueckhahn, 2021).

Ger districts, which take up an estimated 22% of households in the capital (NSO, 2020), present substantial socio-economic challenges. Residents lack essential services such as running water, sewage systems, and central heating (NSO, 2024). Consequently, they rely on coal-burning stoves for heating and cooking, exacerbating air pollution and related health problems (Allen et al., 2013). Moreover, these settlements are not officially recorded and hence limits their access to urban planning efforts, education, healthcare, and transportation services (Hamiduddin and Plueckhahn, 2021). High unemployment and poverty rates in ger districts further deepen economic disparities (see Choi and Enkhbat, 2020: 53-54). These challenges underscore the need of monitoring densely populated ger settlements as a distinct form of urban slums in Mongolia (Boldbaatar et al., 2024; Korytnyi et al., 2023).

Despite gers being emblematic of urban poverty in Mongolia, our understanding of their spatial patterns remains limited. Housing censuses have attempted to assess ger conditions, but traditional methods like surveys require considerable time and cost for accurate measurement. Additionally, the number of gers fluctuates significantly each year due to new migrants from surrounding provinces and evictions resulting from governmental redevelopment plans (Choi and Enkhbat, 2020).

Several previous studies manually digitized the boundaries of ger dwellings from satellite images (Saizen & Tsutsumida, 2017; Tsutsumida et al., 2015). However, this manual approach is impractical for covering the entire region of Ulaanbaatar. Other studies have employed object-based image analysis on satellite images to track changes in ger districts. For example, Park et al. (2019) found that the total area of ger districts expanded from 32 km$^2$ to 221 km$^2$ between 1990 and 2013. However, their use of low-resolution images (i.e., 30-meter resolution) and limited pixel accuracy (i.e., 0.75) hindered precise counting of gers.

In contrast, we employ a machine learning approach with very-high-resolution (VHR) satellite images—with a resolution of less than one meter per pixel—to detect gers in Ulaanbaatar with greater precision. Such AI-based method significantly enhances spatial and temporal coverage by enabling consistent and repeated measurements (Ahn et al., 2023). We collected VHR images over a span of seven years, from 2015 to 2023. Our results estimate the ratio of the slum population in Mongolia, following the criteria of urban slums defined by UN-HABITAT (2021). Considering Ulaanbaatar accounts for 63% of Mongolia's total urban population, it serves as a representative indicator for extrapolating nationwide slum ratio.

This article advances poverty mapping in Mongolia by classifying ger districts as a new form of urban slums based on their socio-economic deprivations. In addition, we demonstrate the effectiveness of AI-based methods for tracking informal settlements over time. Our main findings provide a comprehensive analysis of ger trends between 2015 and 2023, focusing on the persistence and expansion of these districts. Finally, our national-level extrapolation on slum ratio supplements the current understanding of poverty trends in Mongolia, particularly in the aftermath of the COVID-19 pandemic.

2.  Assessing gers in Ulaanbaatar as urban slums

In 2002, UN-Habitat convened an expert group meeting to develop an operational definition of slums. The definition classified a 'slum household' as one experiencing one or more of the following deprivations: (1) lack of access to improved water sources, (2) inadequate living space, (3) poor housing durability, (4) lack of access to improved sanitation facilities, and (5) insecurity of tenure. Subsequent efforts by the United Nations (2007) clarified the sub-definitions, guidelines, and methodologies for each deprivation.

Firstly, an improved water source includes piped water, public taps, and bottled water (UN-HABITAT, 2021:4). Protected springs and wells are also considered acceptable, although their use in urban areas is concerning due to the high risk of contamination (UN-HABITAT, 2003:12). According to the 2020 Population and Housing Census of Mongolia (NSO, 2020), most ger households in urban areas have an improved drinking water supply, including protected wells/springs (83.6%) and piped water (15.4%) (Table 1). Only one percent can be classified as slum households based on this water-related indicator.

Table 1. Housing conditions in Gers, 2020

| Indicators | 2020 Population and Housing Census | | Slum conditions |
|---|---|---|---|
| Drinking water source | Protected wells and springs | 83.6% | X |
| | Kiosk connected with central system | 9.4% | X |
| | Kiosk not connected with central system | 6.0% | X |
| | Unprotected wells and springs | 0.4% | O |
| | Others (e.g., surface water) | 0.6% | O |
| Number of household members in a ger | One or two | 34.6% | X |
| | Three or four | 36.3% | X |
| | Five or six | 23.9% | X |
| | More than seven | 5.2% | X |
| Toilet type | Improved pit latrine with slab | 2.2% | X |
| | Bio-latrine | 0.2% | X |
| | Regular pit latrine with slab | 96.8% | O (in the Mongolian context) |
| | Open pit | 0.8% | O |

Source: NSO (2020: 171, 296).

A sufficient living space is defined as a condition where no more than three household members share a room of at least nine square meters (UN-HABITAT, 2021:4). A ger, which typically consists of a single undivided room with about 30 square meters (Silkroadyurts, 2024), would need to house more than ten residents to be classified a slum based on this criterion. Since most ger households have fewer than seven residents (Table 1), gers cannot be classified as slums based on their residential space. In addition, a house is considered as 'durable', thereby not as slums, when its structure protects inhabitants from the extreme climatic events (UN-HABITAT, 2021:5). A ger, rooted in traditional nomadic knowledge, meets the durability criterion as it provides protection even during *Dzud*—an extreme winter storm in Mongolia where temperatures can drop to minus 46 degrees Celsius (Hahn, 2017).

Regarding improved sanitation, a household is classified as a slum household if its members lack access to a facility that properly separates feces from human contact (UN-HABITAT, 2003: 234). Improved facilities include flush/pour-flush toilets or latrines connected to a sewer, septic tank or pit; ventilated improved pit latrine; pit latrine with a slab or platform, which covers the pit entirely; and composting toilets/latrines (UN-HABITAT, 2021:4). According to this classification, simple pit latrines, which are the most common type of toilet in gers (96.8%; Table 1), can be considered as improved facilities.

However, in the Mongolian context, simple pit latrines cannot adequately separate human waste from residents in ger districts. Although these latrines are occasionally emptied by vacuum trucks during the summer, cleaning cannot be serviced during the winter due to extremely low temperatures (Uddin et al., 2015: 457). When a pit fills up, ger households typically dig a new pit and move the latrine structure, leaving the fecal matter in the backyard where it can infiltrate into the soil and groundwater (Sigel et al., 2012). Such practice, combined with unsafe water storage, heightens the risk of transmitting biological pathogens (Uddin et al., 2014; Oyunbat et al., 2022). Léo Heller, the UN Special Rapporteur on Human Rights, has noted that most residents in urban ger areas lack adequate access to toilets (OHCHR, 2018). Given these conditions, up to 97.6% of urban ger households can be classified as slum households (Table 1).

Lastly, we classify ger districts as slums based on the insecure tenure. Gers are often subject to redevelopment projects and face the risk of forced eviction (Amnesty International, 2015; 2017; 2019). In 2013, the Mongolian government introduced the 'Ulaanbaatar 2020 Master Plan and Amendments and Development Approaches for 2030 (Master Plan)' to address the challenges of unplanned urbanization. One element of this plan is the redevelopment of ger districts, encompassing 75 ger blocks or 1,500 hectares in Ulaanbaatar (M.A.D. Investment Solutions, 2015). The plan identified the need for redevelopment due to inadequate urban public services, poor construction quality, and air pollution (City of Ulaanbaatar, 2014:17). Supporting the plan is the Urban Redevelopment Law, enacted in June 2015. This law authorizes private developers to purchase land from ger residents and carry out development projects. However, individuals without legal land titles are excluded from requesting agreements or information about the projects from the developers. Moreover, the lack of safeguard regulations puts those affected by the redevelopment projects at risk of forced evictions (Amnesty International, 2015).

3. Materials and methods

We gathered multi-temporal satellite images of Ulaanbaatar at a zoom-level 18 resolution. The zoom level system refers to the number of 256-by-256 pixel images required to display the world. These images, called tiles, are organized into a pyramid structure with multiple levels, known as zoom levels. At the apex of the pyramid (i.e., zoom level 0), the entire world is shown on a single tile. Each subsequent zoom level doubles the number of tiles horizontally and vertically, with zoom level 1 displaying the word with 2 X 2 tiles, and zoom level n with $2^n$X $2^n$ tiles. Higher zoom levels display smaller areas with the same number of pixels, thereby revealing more detailed information with higher spatial resolutions. The resolution of tiles varies by latitude because tiles occupy a smaller area as latitude increases. In Mongolia, the resolution of zoom level 18 tiles is approximately 40 centimeters per pixel, and each tile covers an area of 0.01 km$^2$. This resolution is sufficient to clearly identify the boundaries and the roof texture of ger structures (see Figure 2 top).

The study downloaded real-color satellite images with three visible bands (reg, green, and blue) via the ESRI ArcGIS REST API from the World Imagery dataset. The dataset contains images captured at seven temporal periods, covering the downtown area of Ulaanbaatar. The periods include 2015 July, 2016 August, 2019 May, 2020 May, 2021 April, 2022 April, and 2023 June. The yellow boundary in Figure 1 indicates the area where zoom level 18 satellite images exist for all seven periods. This area encompasses 97.1%[1] of the built-up area across six sub-districts of Ulaanbaatar: Bayangol, Bayanzurkh, Songinokhairkhan, Sukhbaatar, Khan-Uul, and Chingeltei. These six sub-districts had a population of 1,567,195 in 2023, representing 63.2% of the total urban population in Mongolia (NSO, 2024). A total of 50,630 tiles were downloaded for each temporal period, amounting to 354,410 tiles across the seven periods.

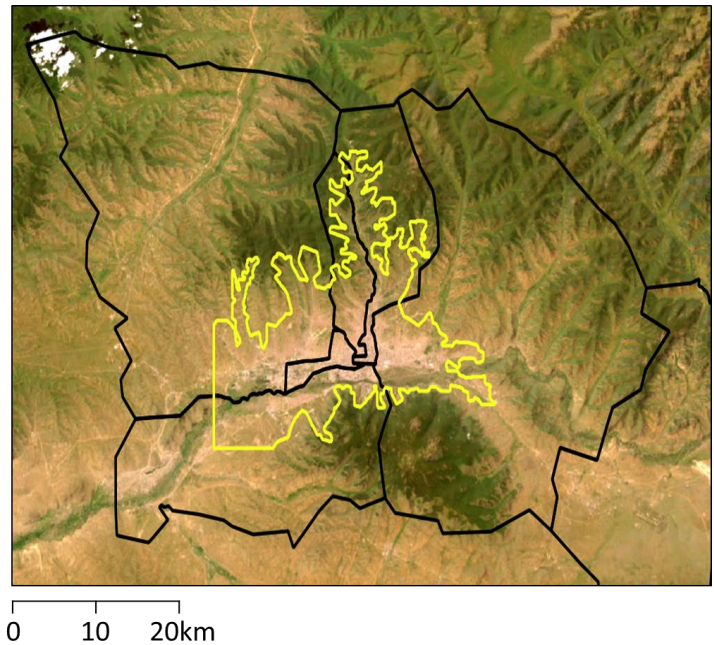

Figure 1. Satellite images used for the study (highlighted within a yellow boundary) and administrative boundaries of Ulaanbaatar's sub-districts (black). The background images are sourced from World Imagery.

To train a machine learning model to detect gers from satellite images, we created label data due to the lack of ground-truth data on the spatial location of gers. Since the location of a ger can shift subtly over time, a new label must be created for each different temporal image, even if it was taken at the same location. Labels were created for satellite images taken in 2022 and 2016. Using QGIS 3.14, we manually digitized the footprints of 7,305 gers, visually confirmed from 2,002 images (Figure 2, red circles). The ger's clearly distinguishable appearance and consistent size in the satellite images facilitated the digitization process, which was completed in one week. The digitized footprints were then transformed into images of 256-by-256 pixel images, matching the size of satellite images. Pixels containing gers were assigned a value of 1 and those representing non-gers were assigned a value of 0. The labels can be confirmed through uploaded Research Data.

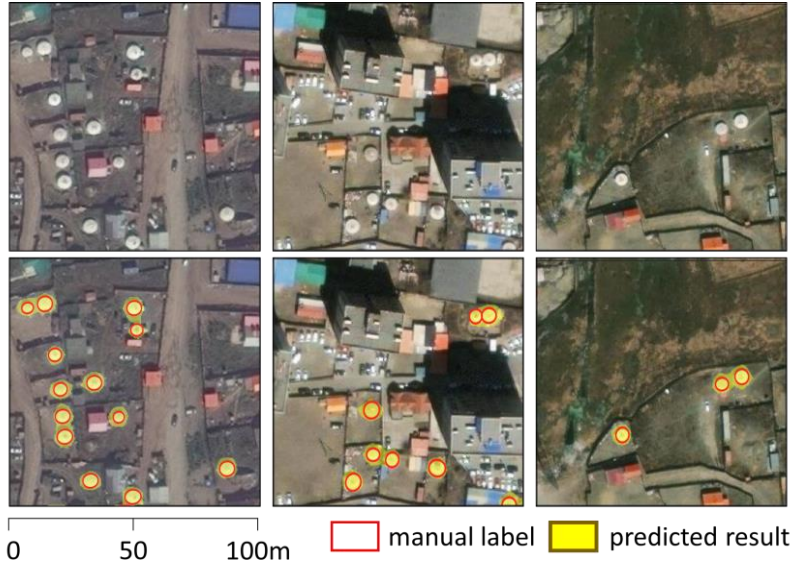

Figure 2. Examples of input satellite images, manual labels (red), and predicted results from Deep Lab v3 model (yellow). The background images are sourced from World Imagery.

The task of our segmentation model is to predict $y \in \{0,1\}^{H \times Y}$, the binary pixel value of 0 (non-gers) or 1 (gers), based on $x \in \mathbb{R}^{H \times W \times 3}$, the pixel value of the input satellite images. The DeepLabv3 (Chen et al., 2017) model was used for slum segmentation in a supervised manner. DeepLabv3 is a segmentation model that utilizes atrous spatial pyramid pooling (ASPP) layers to capture features at multiple scales. ASPP consists of multiple parallel atrous convolution, which incorporates a dilation rate into standard convolution. In the context of a 2-dimensional signal, given a convolution filter $\omega$, an atrous rate $r$, and an input feature map $m$, the output feature map $n$ resulting from the application of atrous convolution is expressed as follows:

$$n[i] = \sum_k m[i + r \times k] \ \omega[k] \qquad (1)$$

Additionally, the slum dataset is highly imbalanced, with a very small proportion of slum instances. To address this imbalance during training, focal loss (Lin et al., 2017) was employed. Focal loss is a modified version of cross entropy loss designed for imbalanced datasets. Cross entropy loss can be mathematically expressed as follows:

$$CELOSS = -\frac{1}{H \times W} \sum_{i=0}^{H \times W} y \cdot \text{LogSoftmax}(\hat{y}_i) \qquad (2)$$

where $\hat{y}_i$ means the model prediction for pixel $i$.

Then, the focal loss is:

$$\alpha = \sum_{c=1}^{C} \theta_C \times \mathbb{1}[\text{argmax}(\hat{y}) = c] \qquad (3)$$

$$FocalLoss = CELoss \times (1 - exp(-CELoss)) \times \alpha \qquad (4)$$

where $\hat{y}$ is a target prediction and $\theta_c$ is a weight of class $c$. We used $\theta_c = (0.1, 0.9)$ for non-slum and slum. ResNet-50 was utilized as the backbone and 1800 manual labels (around 90%) are used in training.

The model's performance was evaluated using the remaining 10% of manual labels. The assessment metrics included the F1 score, mean Intersection over Union (i.e., mIoU), and pixel accuracy for binary classifications of ger and non-ger. The F1 score, which balances precision and recall, measures the model's accuracy in identifying gers. The mIoU evaluates the overlap between predicted and actual regions, providing insight into the model's localization performance. Pixel accuracy measures the proportion of correctly classified pixels. The evaluation results for ger and non-ger categories are as follows: F1 scores of 0.65 and 0.99, mIoU values of 0.48 and 0.99, and pixel accuracies of 1.00 and 0.99, respectively. For semantic segmentation models, F1 Score or mIoU are more commonly used as performance evaluation indicators than pixel accuracies. An F1 score of 0.65 and mIoU of 0.48 for gers indicate a moderate level of performance. The predicted results evaluated an area larger than the actual ger area as a ger (see Figure 2 below). However, since the goal of this study is not to detect the exact size and boundaries of gers, but rather to determine the number of gers, this level of performance is deemed sufficient.

Since the model's results show the area of the gers rather than generating the number of gers directly, the number must be calculated from the detected area. The average area of a ger was found to be 61 square meters based on the model's detection. To estimate the number of gers, the total detected area was divided by 61 and then rounded to the nearest whole number.

The final step involved human verification of the results. During this process, other structures with circular shapes and white roofs (e.g., observatories, circular dome buildings) that were also detected were manually removed. Additionally, in cases where multiple gers overlapped and were detected as a single object, the number of gers was directly counted and corrected. This human inspection took a total of two days. Detection results before and after verification can be confirmed through uploaded Research Data.

## 4. Results

Yellow dots in Figure 3 shows individual ger structures detected in 2023. The figure (right) confirms that gers are relatively absent in the center of Ulaanbaatar, where high-rise buildings are concentrated. Instead, gers are predominantly found in the peripheral areas surrounding the city center, and they tend to spread out to the north along mountain ranges and riversides. The reason gers spread to the north is due to the presence of the Tuul Gol River and Bogd Khan Mountain to the south of downtown, which limits the availability of flat land for gers in the southern area.

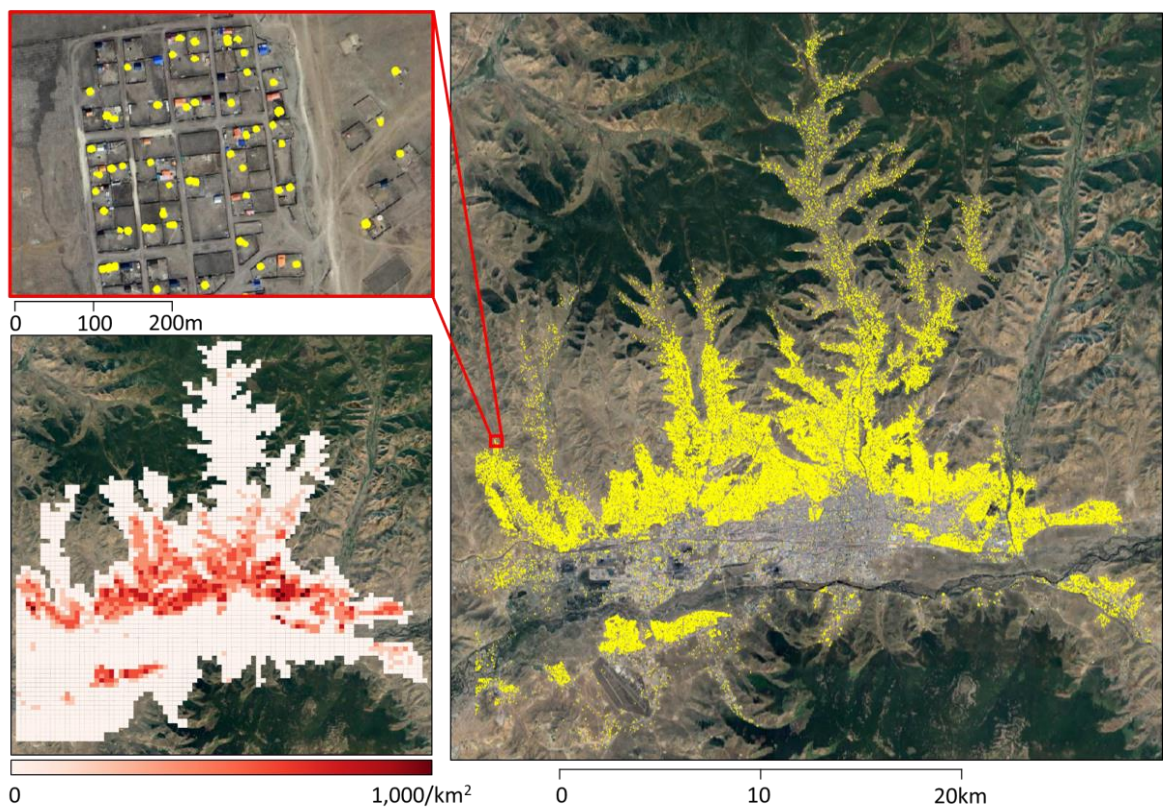

Figure 3. (Right) The detected gers in Ulaanbaatar from satellite images of 2023. (Top-left) The zoomed-in views of the result. (Bottom-left) Ger density map per square kilometer. The background images are sourced from World Imagery and Google Maps.

The zoomed-in result (Figure 3 top-left) shows the typical layout of ger settlements in peri-urban areas of Ulaanbaatar. These settlements consist of square-shaped wooden fences, called *khashaas*, surrounding individual gers and the land within. Khashaas are used to delineate property boundaries for ger residents (Dore and Nagpal, 2006). There are also makeshift wooden houses identified in satellite images and, increasingly, more permanent brick or concrete structures within khashaas as residents seek better living conditions. The settlement is connected to the city center by unpaved roads, leading to muddy conditions during the rainy season and dust in the dry season, as shown in the eastern part of the figure.

The gray boxes in Figure 4 illustrate the change in the number of gers in Ulaanbaatar over the past decade, revealing three major trends. The most prominent trend is the sharp decline in gers from 2016 to 2019, with a decrease of 10,492. This decline is likely due to the ger redevelopment projects initiated as part of Ulaanbaatar 2020 Master Plan, which began in 2013. Although attempts to redevelop ger areas by the private sector have existed since 2008 (M.A.D. Investment Solutions, 2015), the projects gained official support from the Mongolian government following the legal regulation enacted in 2015 (Choi and Enkhbat, 2020). The land acquisition process for these projects had been delayed and actual development began in 2016, with a target completion in the early 2020s. This timeline corresponds with the rapid decline in gers during that period. In total, 11,755 new apartment complexes were constructed by 43 individual projects on the 2,519 units of land where gers were originally located (Montsame, 2022).

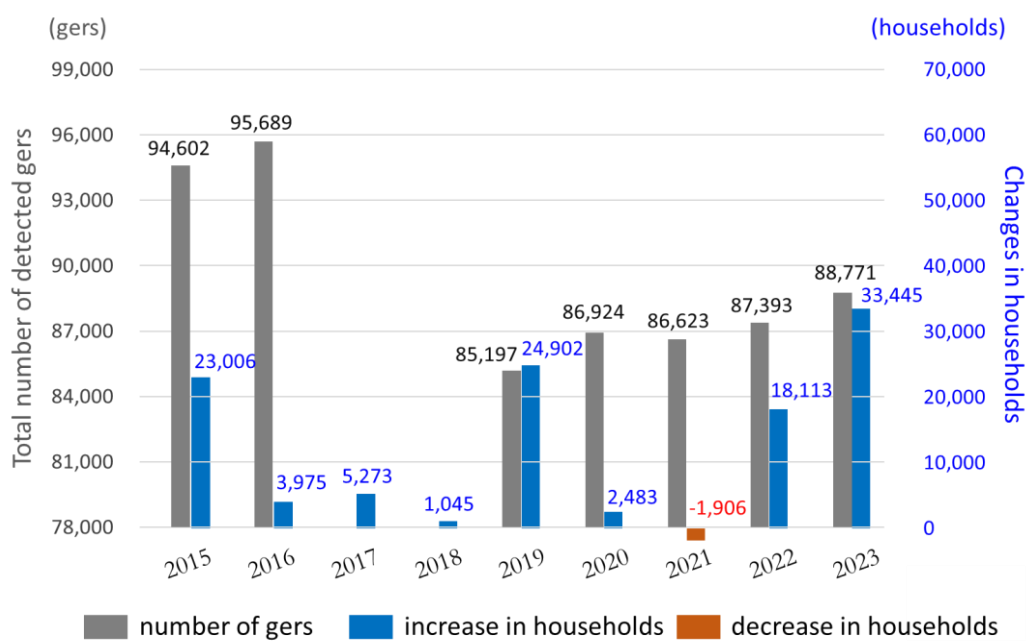

Figure 4. The number of detected gers (gray) and increase/decrease in households (blue/red), 2015 - 2023. Household data of six sub-districts of Ulaanbaatar (Bayangol, Bayanzurkh, Songinokhairkhan, Sukhbaatar, Khan-Uul, and Chingeltei) were sourced from NSO (2024).

The second trend is the sporadic increases of gers in 2015/16, 2019/20, and 2022/23. During these three periods, the ger settlements increased by 1,087, 1,727, and 1,378 respectively. This increase appears to be correlated with a rise in total households within the six sub-districts.[2] During the same periods, the total number of households increased by approximately 20,000 (shown in blue boxes in Figure 4). Given that in-migration from surrounding provinces has been a dominant factor in Ulaanbaatar's population growth over the last three decades (Otgonkhuu et al., 2023), the increase in gers can be attributed to rural migrants bringing their own gers. If each newly established ger accommodates one household, it can be estimated that about 5-7% of the increased households reside in gers. In 2019, when the number of new households increased by 24,902, and approximately 6.9% of them (i.e., 1,727 households) are likely to have moved into gers.

Lastly, the number of gers remained stable at around 87,000 from 2020 to 2022. This stability can be attributed to the impact of the global COVID-19 pandemic. In February 2020,

<hr />

[2] Note that household statistics were adjusted by one year to align with the temporal periods of satellite images. Satellite images were taken at the midpoint of the year (e.g., April 2022), while household statistics are based on data compiled at the end of the year. For instance, the household statistics for 2022 reflect data compiled in December 2022, resulting in a timing discrepancy. Consequently, the 2022 satellite image data was matched with the 2021 household statistics.

the Mongolian government began restricting all travel between provinces and cities, as well as incoming and outgoing flows in Ulaanbaatar (IOM, 2020). Multiple lockdowns were imposed in the capital (Krusekopf & Jargalsaikhan, 2021) to prevent it from becoming a nationwide source of disease transmission. Restrictions on movement were gradually lifted starting in April 2021, followed by a complete life in July 2021 (Dorjdagva et al., 2022). The stable population during the period reflects these restrictions, highlighting rural-to-urban migration as a major driver of Ulaanbaatar's population growth. Notably, the number of households even decreased by 1,906 in 2021 (a red box in Figure 4). Since the stable population, the number of gers also remained stable during this period.

## 5. Discussion

### 5.1. Extending to periphery of Ulaanbaatar

Our findings demonstrate the powerful application of AI and satellite data in analyzing the geospatial marginalization of ger settlements. Figure 5 depicts the grid-level changes in prominent ger locations from 2015 to 2023. Each grid covers an area of 0.67 square kilometers and is in zoom level 15; a grid further consists of 64 tiles that are a higher visual resolution of zoom level 18. The grid colors represent changes in ger counts, with red denoting an increase and blue a decreasing. The figure shows that ger locations have moved from the center to the periphery of the city, mirroring the urban sprawl phenomenon.

The figure also highlights locations of 19 yellow dots corresponding to the redevelopment projects in the area (M.A.D. Investment solutions, 2015:40). Satellite images reveal a trend where ger settlements near these project sites decreases as new high-rise buildings emerge. Nonetheless, ger counts increase in the surrounding grids. Particularly, in the western and eastern parts of the city center, many grids are depicted in deep red, indicating a significant increase in ger counts. This trend suggests that ger residents are being relocated to the outskirts of Ulaanbaatar due to the government's redevelopment policy. It also indicates that the benefits of redevelopment were not equitably shared with the displaced ger residents (Choi and Enkhbat, 2020), and new policy decisions could further help address socio-economic disparities among urban residents.

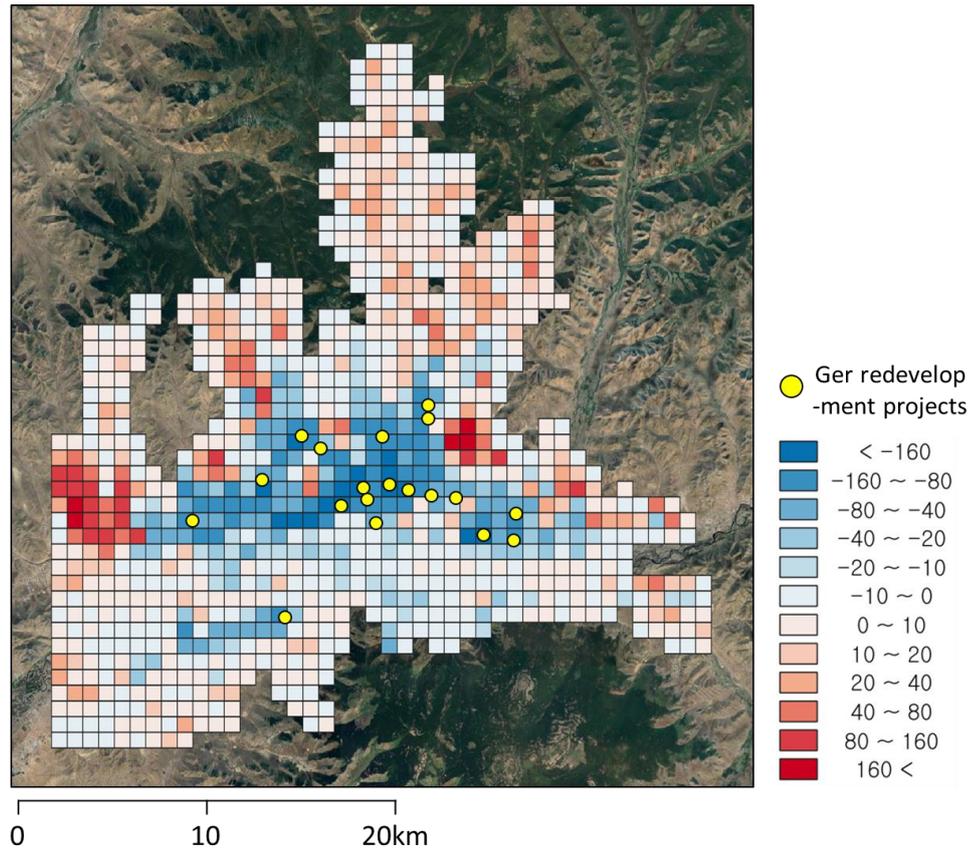

Figure 5. Changes in ger counts from 2015 to 2023. The locations of ger redevelopment projects (yellow dots) were sourced from M.A.D. Investment Solutions (2015:40).

Our results also suggest that the living conditions of ger residents have been increasingly marginalized since 2015, particularly in terms of accessibility. We can confirm this trend from another data, as shown in Table 2. Utilizing the global high spatial resolution digital elevation model (GDEM) version 3 from ASTER (Advanced Spaceborne Thermal Emission and Reflection Radiometer; Abrams et al., 2022), we assessed the average elevation and slope of ger settlements in 2015 and 2023. The table shows increases in both elevation and slope over the decade, indicating that the gers have progressively relocated to highly elevated areas on the outskirts of the city.

Shifted locations also mean that accessibility to critical infrastructure and service becomes limited for ger residents. To examine this, we measured the average distance from gers to major facilities within Ulaanbaatar, based on the spatial coordinates of facilities from OpenStreetMap (2024), which is a collaborative project that provides free geographic datasets. This dataset includes locations of points, lines, and areas of interest, classified into various categories. Specifically, we collected information on main roads, bus stations, educational institutions, medical facilities, public amenities, marketplaces, and industrial zones, utilizing the classifications available within OpenStreetMap. For example, the main roads were categorized under classes such as "primary," "secondary," "tertiary," and "trunk." Finally, we calculated the average distance to Sükhbaatar Square, a central and significant public space known for hosting cultural events, ceremonies, and gatherings.

The calculation of these average distances indicated that, for all categories, the distance from gers to critical services increased by 9 to 24% in 2023 compared to 2015 (Table 2). For instance, as gers have shifted to the peripheral areas, residents now live approximately 200 meters farther from public amenities, with the distance from 1,970 to 2,160 meters. These findings are consistent with previous qualitative studies (e.g., Hamiduddin and Plueckhahn, 2021), which suggest that residents of gers face lower accessibility compared to those living in the central areas of the capital.

Table 2. Changes in geographical indicators of gers between 2015 and 2023

| Indicators | 2015 | 2023 | class |
|---|---|---|---|
| Elevation (meter) | 1,365.9 | 1,371.1 | |
| Inclination (degree) | 2.59 | 2.64 | |
| Distance to city center (meter) | 8,059.7 | 8,789.7 | Sükhbaatar Square |
| Distance to main roads (meter) | 1,162.7 | 1,319.5 | primary, secondary, tertiary, trunk |
| Distance to bus stations (meter) | 857.7 | 1,034.2 | bus station, bus stop |
| Distance to education institutions (meter) | 1,928.9 | 2,373.9 | college, school, university |
| Distance to medical facilities (meter) | 1,202.6 | 1,446.5 | dentist, doctors, hospital, pharmacy |
| Distance to public amenities (meter) | 1,970.3 | 2,160.0 | community center, fire station, police, public building, post office, town hall |
| Distance to marketplaces (meter) | 265.5 | 330.2 | department store, convenience, mall, market place, supermarket |
| Distance to industrial zones (meter) | 2,116.6 | 2,450.8 | industrial |

## 5.2. Estimating slum population ratio based on ger detection

We estimate the slum population ratio in Mongolia based on the number of gers detected in satellite images. To calculate the ratio, we first divide the number of households living in gers by the number of years of gers detected. According to the 2020 Population and Housing Census (NSO, 2020), there were 91,249 households living in gers in six sub-districts of Ulaanbaatar, while 86,925 gers were detected from satellite images from the same year. This results in an average of 1.026 households per ger. Using this average, we can estimate the number of households for other years by multiplying 1.026 by the number of gers detected for those years. We assumed that a consistent number of households per ger during this time. For the ger-dwelling ratio, we divide the number of households living in gers by the total number of households in the area. This ratio serves as a proxy for poverty levels and the slum population in Mongolia.

For validation, we compare this proxy with district-level poverty headcount data. The World Bank report (Singh et al., 2017) calculated poverty levels at the district level using small area estimation techniques based on the 2010 census and the 2011 household socio-economic survey. This report provides the detailed percentage of people living in poverty in 2011 across 23 districts of Ulaanbaatar. Our analysis estimated the ger-dwelling ratio by districts in 2015, revealing a strong correlation of 0.836 with the 2011 poverty statistics. This significant correlation suggests that gers are predominantly situated in regions with higher urban poverty, thus supporting the use of AI-based ger-dwelling ratio as a reliable proxy for poverty (see Supplementary Materials A).

The national-level ger-dwelling ratio is depicted in Figure 6 (red) alongside the estimates from UN-HABITAT (blue). According to UN-HABITAT (2024), which updates its estimates biennially, there has been a gradual decline in the slum population ratio in Mongolia. However, several critics (Carr-Hill, 2013; Breuer et al., 2024; Lucci et al., 2018) have suggested that these estimates may be understated. Additionally, UN-HABITAT's estimates have not been updated since 2020, leaving the effects of the COVID-19 pandemic on slum conditions unknown. In contrast, our estimates show greater variability. Between 2015 and 2020, the proportion of slums is expected to decrease in our estimates at a slower rate than UN-Habitat's estimates. More critically, our findings indicate that since 2020, including the pandemic period, the slum rate has stabilized at 21% of the urban population and our estimate is that this number continues to hold for 2023 (i.e., 21.1%). This suggests that urban poverty in Mongolia has persisted through the pandemic.

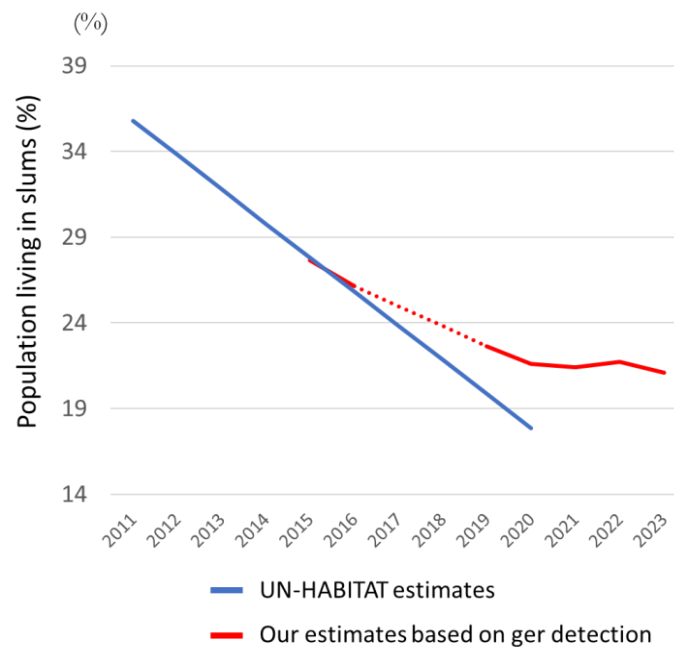

Figure 6. Population living in slums in Mongolia (% of urban population). The blue line represents estimates from UN-HABITAT (2024), while the red line shows our estimates based on the detection of gers (2015 - 2023).

It should be noted that our estimates may be conservative, and the actual slum ratio may be higher. This is partly because the ger-dwelling population ratio in urban areas is higher in areas outside of Ulaanbaatar compared to the six sub-districts used in our analysis (NSO, 2024). Within these six sub-districts, 21.6% of households reside in gers. There are 17 other sub-districts in Mongolia where all households are classified as urban households, similar to the six sub-districts, and their ratio is 33.0% (see Supplementary Materials B). Therefore, incorporating these higher ger-dwelling ratios could reveal a more severe situation than what is currently reflected in our findings.

Finally, our study considered only households living in gers as representing the slum population in Mongolia. This excludes other potential slum types, such as shanty towns constructed from scrap materials, rooming houses that rent out individual rooms to multiple occupants, refugee camps, and informal settlements (see Davis, 2006). Additionally, even if an apartment is constructed with durable materials, it could still be classified as a slum if it lacks running water and sanitation facilities. However, satellite imagery cannot be used to detect the interior conditions of buildings, making it impossible to calculate the proportion of such slums. Therefore, including these other forms of slum settlements would likely result in a higher slum ratio than our current estimates.

## 6. Conclusion

In this paper, we presented an AI-based approach to analyzing the spatial and temporal dynamics of ger settlements in Ulaanbaatar, Mongolia. Utilizing VHR satellite imagery, our model tracked the evolution of ger settlements from 2015 to 2023. Our findings indicate a significant reduction in the number of gers due to urban redevelopment projects, coupled with a notable increase in ger dwellings in the city's peripheral areas. Such displacement underscores the marginalization of ger residents, who now face longer distances to essential infrastructure (Table 2).

Using the ger detection results, we extrapolated the nationwide slum ratio in Mongolia. The extrapolation confirms a decline in the slum population from 2015 to 2020, consistent with official reports (UN-HABITAT, 2024). However, from 2020 onwards, the slum ratio stabilizes at approximately one-fifth of the population. This stabilization suggests that urban poverty since COVID-19 may be more severe than what is reflected in official reports published by the Mongolian government and UN-Habitat (Figure 6).

This work classified densely populated urban ger settlements as slums due to their poor living conditions, characterized by inadequate sanitation and insecure tenure. These gers have lost their mobility, being permanently settled in specific areas with fencing (Tsutsumida et al., 2015). In contrast, labeling gers of the Mongolian steppes as slums is politically sensitive, given their role as traditional dwellings for nomadic communities. Moreover, the gers constructed for international tourists in Mongolia's natural parks outside Ulaanbaatar (e.g., Gorkhi Terelj National Park) do not meet the criteria for slums, as they are equipped with modern amenities and have close proximity to nearby resorts.

A significant factor contributing to the sustained slum ratio is the continuous migration of nomadic families from the Mongolian steppes, who bring their own gers and begin to settle in urban areas (NSO, 2024). This migratory trend is often driven by the search for better employment opportunities and to avoid environmental hazards such as *dzud* (Xu et al., 2021). Dzud causes massive livestock fatalities, devastating the livelihoods of nomadic communities and accelerating their migration to cities (Algaa, 2023). The frequency and intensity of dzud are expected to rise in the future due to global environmental change (Fernández-Giménez et al., 2012), potentially exacerbating the influx of migrants into ger districts in Ulaanbaatar.

Our findings carry significant implications for the United Nations' Sustainable Development Goals (SDGs) concerning urban slums. Recognizing slums as a critical urban challenge, the UN has identified the slum population ratio as one of key indicators for 'Sustainable Cities and Communities', the 11th SDG (Indicator 11.1). According to UN-HABITAT (2024), the international community was on track to meet this target by 2020 across nearly every continent. However, our evidence from Mongolia suggests that progress may be slower than anticipated. Moreover, trends in urban slums and poverty appear to have persisted in the aftermath of the COVID-19 pandemic, consistent with findings from previous studies (Kansiime et al., 2021; Sumner et al., 2020). These findings call for the urgent need for renewed efforts and innovative strategies to address urban poverty, inclusive of ger residents.

**Declaration of interest statement:**

The authors declare that there are no competing interests that could influence the work reported in this paper.

**Declaration of Generative AI in Scientific Writing**

During the preparation of this work the author(s) used ChatGPT in order to proofread the draft. After using ChatGPT, the authors reviewed and edited the content as needed and take full responsibility for the content of the published article.

**Data availability**

The labels used for training the model and the ger detection results have been uploaded. However, the satellite image data has not been uploaded due to the copyright restrictions of World Imagery.

**Acknowledgements:**

The authors thank Soyoung Lee, Donghyun Ahn, Hyunjoo Yang, Sangyoon Park, Jihee Kim, and colleagues from the Institute for Basic Science for their valuable comments on the initial conceptualization of this paper.

**CRediT authorship contribution statement**

**Jeasurk Yang:** Conceptualization, Data curation, Investigation, Resources, Visualization, Writing – original draft, Writing – review and editing.

**Sumin Lee:** Formal analysis, Methodology, Software, Validation, Writing – original draft,

**Sungwon Park:** Investigation, Methodology, Writing – original draft

**Minjun Lee:** Formal analysis, Methodology, Software, Validation

**Meeyoung Cha:** Funding acquisition, Project administration, Supervision, Writing – review and editing.